\pdfoutput=1
\RequirePackage[T1]{fontenc}
\documentclass[12pt]{article}

\usepackage[height=8.85in,width=6.45in]{geometry}

\usepackage[utf8]{inputenc}
\usepackage{amsmath}
\usepackage{amssymb}
\usepackage{mathtools}
\numberwithin{equation}{section}
\usepackage{slashed}
\usepackage{braket}
\usepackage[svgnames]{xcolor}
\usepackage[colorlinks,citecolor=DarkGreen,linkcolor=FireBrick,urlcolor=FireBrick,linktocpage,unicode]{hyperref}
\usepackage{cite}
\usepackage{graphicx}
\usepackage{tikz}
\tikzset{>=stealth}

\usepackage{times}
\usepackage{courier}
\usepackage{bm}
\usepackage{subfig}

\usepackage{xcolor}
\usepackage{mdframed}

\renewenvironment{figure}[1][]{
  \begin{originalfigure}[#1]
    \begin{mdframed}[linecolor=black!0,backgroundcolor=black!1]
}{
    \end{mdframed}
  \end{originalfigure}
}

\usepackage{mathrsfs}

\def\Nequals#1{$\mathcal{N}{=}#1$}
\def\pt{\mathrm{pt}}
\def\bZ{\mathbb{Z}}
\def\bR{\mathbb{R}}
\def\bC{\mathbb{C}}
\def\cH{\mathcal{H}}
\def\cL{\mathcal{L}}

\def\TMF{\mathrm{TMF}}
\def\SQFT{\mathrm{SQFT}}

\def\KO{\mathrm{KO}}
\def\tr{\mathop{\mathrm{tr}}}

\def\vev#1{\langle#1\rangle}

\def\Coker{\mathop{\mathrm{Coker}}}
\def\Ker{\mathop{\mathrm{Ker}}}
\def\Im{\mathop{\mathrm{Im}}}
\def\forget{F}

\begin{document}

\begin{titlepage}

\begin{flushright}
\end{flushright}

\vskip 3cm

\begin{center}

{\Large \bfseries 
On  a long exact sequence of groups of\\[.5em]
equivalence classes of 2d \Nequals{(0,1)} SQFTs
}

\vskip 1cm
 Yuji Tachikawa
\vskip 1cm

\begin{tabular}{ll}
  &Kavli Institute for the Physics and Mathematics of the Universe, \\
 & University of Tokyo,  Kashiwa, Chiba 277-8583, Japan
\end{tabular}

\vskip 2cm

\end{center}

\noindent

Strongly motivated by a mathematical result by Lin and Yamashita \cite{LinYamashita},
we describe a long exact sequence formed by 
groups of equivalence classes of 
two-dimensional \Nequals{(0,1)} supersymmetric quantum field theories (SQFTs)
with and without $SU(2)$ symmetry.
As an application, we study chiral fermions
in heterotic compactifications 
with $SU(2)$  symmetry of level one to four dimensions,
and show that each even-dimensional irreducible representation of $SU(2)$ appears 
even times,
assuming the conjectural relation between topological modular forms and SQFTs.
This implies the absence of the Witten anomaly, but contains more information than that.

\end{titlepage}

\setcounter{tocdepth}{2}
\tableofcontents

\section{Introduction and summary}
\label{sec:introduction}

\paragraph{Equivalence class of theories.}
There is a growing list of evidence 
that it is quite fruitful to introduce a certain equivalence relation among 
two-dimensional (2d)   \Nequals{(0,1)} supersymmetric unitary quantum field theories (SQFTs),
and that the set of equivalence classes under this equivalence relation equals what is known in mathematics as topological modular forms (TMFs).
Here, the equivalence relation is obtained by, roughly speaking,
identifying two 2d \Nequals{(0,1)} SQFTs 
continuously connected by supersymmetric deformations and renormalization group flows,
and regarding as zero any such SQFT breaking supersymmetry spontaneously.

The origin of this line of thought ultimately goes back to a foundational paper by  \cite{Witten:1986bf} in the 1980s,
but the modern developments started with the conjecture by two mathematicians 
Stolz and Teichner in this century \cite{StolzTeichner1,StolzTeichner2}.
Physicists started to take up this idea again about a decade ago, and in the last few years 
we have seen a flow of ideas in both directions between mathematics and physics.%
\footnote{%
For a partial list of related papers, see \cite{Gaiotto:2018ypj,Gukov:2018iiq,
Gaiotto:2019asa,Gaiotto:2019gef,
Johnson-Freyd:2020itv,Lin:2021bcp,Lin:2022wpx,Albert:2022gcs,
Tachikawa:2023nne,Kaidi:2023tqo,Tachikawa:2023lwf,
Tachikawa:2024ucm,Saxena:2024eil,Kaidi:2024cbx,TachikawaYonekura}.}
For example, the absence of global anomalies of all possible heterotic compactifications was discussed in this context in \cite{Tachikawa:2021mvw,Tachikawa:2021mby,Yonekura:2022reu}.

Since most of the developments on the physics side so far concerned equivalence classes of 2d \Nequals{(0,1)} SQFTs
without any additional symmetry,
extending the analysis to the case with symmetry is a natural next step.
On the side of mathematics, this corresponds to the study of twisted equivariant TMFs,
about which there has been a slow but steady progress, e.g.~in \cite{GepnerMeier};
here `equivariant' means `symmetric' and `twisted' means `with anomaly' on the physics side.
Pursuing this direction further, Lin and Yamashita largely determined 
the structure of twisted $U(1)$- and $SU(2)$-equivariant TMFs in a paper \cite{LinYamashita}.
Translated back to physics language, this means that we can now determine 
the structure of equivalence classes of 
2d \Nequals{(0,1)} SQFTs with $U(1)$ or $SU(2)$ symmetry.
The objective of this short paper is to formulate and derive (a tiny subset of) 
the results of \cite{LinYamashita} in  physics language for physicists, 
and to give some easy applications.
For brevity, we simply refer to 2d \Nequals{(0,1)} SQFTs as SQFTs in the following.

To be more concrete, recall that the gravitational anomaly of a 2d fermionic theory
is specified by an integer $n$,
where we take the convention
that $n=2(c_L-c_R)$ when the theory is conformal,
and that \Nequals{(0,1)} supersymmetry is on the right-movers.\footnote{%
\label{foot:conv0}%
In the literature it might be more  common instead to find the opposite convention 
where $\nu=2(c_R-c_L)$ satisfying $\nu=-n$ is used to
label the gravitational anomaly.
The reason we use $n$ in this paper will be explained in footnote~\ref{foot:conv}.
}
We will introduce an equivalence relation among SQFTs,
and consider the set of equivalence classes of SQFTs with gravitational anomaly $n$,
which we denote\footnote{%
It is also common in the literature to use $\SQFT^n$ to denote the `space' of theories 
(which is expected to form an $\Omega$-spectrum)
and $\SQFT^n(\pt)$ to denote the set of equivalence classes of theories.
In this paper, we use $\SQFT^n$ to denote the set of equivalence classes of theories,
for the sake of brevity.
} by $\SQFT^n$.
The set of SQFTs has a natural structure of a ring,
where the addition and the multiplication correspond to the direct sum and the tensor product of the Hilbert spaces of the theories involved.
The addition makes $\SQFT^n$ an Abelian group, 
and the multiplication makes $\bigoplus_n \SQFT^n $ a graded-commutative algebra.

\paragraph{Some equivalence classes.}
To discuss examples, we consider \Nequals{(0,1)} sigma models on  manifolds $M$
equipped with the $B$ fields satisfying appropriate conditions.\footnote{%
One is the requirement 
at the differential form level that
the field strength $H$ of the $B$ field satisfies $dH=p_1/2$, 
where $p_1$ is the first Pontryagin class of $M$.
Other subtler conditions are detailed in \cite{Witten:1985mj,Yonekura:2022reu}.
Mathematically such a choice is known as a (differential) string structure.}
This sigma model then determines a class 
in $\SQFT^{-\dim M}$.
Concretely, let us consider the \Nequals{(0,1)} sigma model on $S^3$
with the field strength $H$ of the $B$-field
given by $\int_{S^3} H=1$.
It determines a class in $\SQFT^{-3}$ which we denote by $\nu$.
In \cite{Gaiotto:2019asa,Gaiotto:2019gef}, it was shown that 
this class $\nu$ generates a subgroup \begin{equation}
\bZ_{24}\subset \SQFT^{-3} 
\label{contain}
\end{equation} 
and that the \Nequals{(0,1)} sigma model on $S^3$ with $\int_{S^3}H=k$ corresponds to the class $k\nu\in \bZ_{24}$.

Recall also that the anomaly of the $SU(2)$ symmetry is again specified by an integer $k$.
We take the convention that this integer $k$ is given,
when the theory is conformal, 
by the difference $k_L-k_R$ of the levels $k_L$ and $k_R$ of  the $SU(2)$ symmetries
on the left-movers and the right-movers, respectively.
By a slight abuse of terminology, we also call this combination $k$ the level of the $SU(2)$ symmetry,
and we sometimes just write $SU(2)_k$ for an $SU(2)$ symmetry of level $k$.

We can now  consider the Abelian group $\SQFT_{SU(2)}^{n,k}$ of equivalence classes of $SU(2)$-symmetric SQFTs
with gravitational anomaly $n$ and $SU(2)$ level $k$.
We note that the multiplication defines a bihomomorphism 
\begin{equation}
\SQFT^{n,k}_{SU(2)} \times \SQFT^{n',k'}_{SU(2)} \to \SQFT^{n+n',k+k'}_{SU(2)}.
\end{equation}
We also note that there is a natural map \begin{equation}
\SQFT_{SU(2)}^{n,k}\xrightarrow{\forget}\SQFT^n
\end{equation}
obtained by forgetting the $SU(2)$ symmetry.

For example, an $SU(2)$ doublet of free  left-moving Weyl fermions
determines a class $\vec\psi\in\SQFT^{4,1}_{SU(2)}$.
As another set of examples, consider again the \Nequals{(0,1)} sigma model on $SU(2)\simeq S^3$ equipped with
a $B$-field whose field strength $H$ has the flux $\int_{S^3}H=k$. 
Assuming that the metric and the $B$-field are both $SU(2)\times SU(2)$ symmetric,
it is an SQFT with $SU(2)_{k-1}\times SU(2)_{-k-1}$ symmetry;
 for a review, see \cite[Sec.~2]{Gaiotto:2019asa}.
This theory then defines a class $\nu(k)\in \SQFT^{-3,k-1}_{SU(2)}$, by forgetting the second $SU(2)$ symmetry.
By further forgetting the first $SU(2)$ symmetry, we have $\forget(\nu(k))=k\nu$.

Mathematicians have a rigorous definition of the group $\TMF^n$ of topological modular forms
and $\TMF^{n,k}_{SU(2)}$ of twisted $SU(2)$-equivariant topological modular forms.
The proposal of Stolz and Teichner \cite{StolzTeichner1,StolzTeichner2},
generalized to include the case with symmetries,
postulates that\footnote{%
\label{foot:conv}%
This is a good place to continue the discussion we started in footnote~\ref{foot:conv0} 
on why we use $n$ rather than $\nu=-n$ as the label for the gravitational anomaly.
For any non-equivariant (co)homology theory $E$, there is a general equality  $E^n=E_\nu$
between the cohomology groups of the point $E^n:=E^n(\pt)$ and the homology groups of the point $E_\nu:=E_\nu(\pt)$, where $\nu=-n$.
In particular, an SQFT whose gravitational anomaly is specified by $n=-\nu$ is expected to 
determine a class in $\TMF^n = \TMF_{\nu}$. 
In this sense, $n$ is the cohomological degree
and $\nu$ is the homological degree of the same $\TMF$ class which would 
correspond to the same SQFT,
and both $n$ and $\nu$ serve our purpose equally well.
However, in the equivariant case, there is no such general equality between generalized homology groups
and generalized cohomology groups,
and an SQFT with a given gravitational anomaly and a level under a symmetry 
is thought to determine a generalized cohomology class rather than a generalized homology class.
This makes  $n$ rather than $\nu$ a more convenient label in our context.
} we have  isomorphisms\footnote{%
One of the most interesting aspects to the author of these conjectural isomorphisms  
is that the current mathematical 
definition of topological modular forms uses elliptic curves defined over $\bZ$
and other commutative algebras, not necessarily over $\bC$,
making it maximally distant (in some sense) 
to the ordinary study of quantum field theories which are done on
two-dimensional surfaces.
}  \begin{equation}
\SQFT^n \simeq \TMF^n,\qquad \SQFT^{n,k}_{SU(2)} \simeq \TMF^{n,k}_{SU(2)}.
\label{conj}
\end{equation}
Mathematicians have a way to construct $\TMF$ classes from string manifolds,
which can be used to define the classes $\nu\in \TMF^{-3}$, $\nu(k)\in \TMF^{-3,k-1}_{SU(2)}$
and $\vec\psi\in \TMF^{4,1}_{SU(2)}$.
Furthermore, it is known that \begin{equation}
\TMF^{-3}\simeq \bZ_{24},
\end{equation} which is compatible with \eqref{contain}.

\paragraph{The long exact sequence.}
In the mathematical study  of twisted equivariant topological modular forms in \cite{LinYamashita},
an important role was played by various types of long exact sequences.
When we translate one such long exact sequence to  physics language using \eqref{conj}, 
we obtain the following:
\begin{equation}
\cdots\xrightarrow{\nu(k)\times} 
\SQFT_{SU(2)}^{n-4,k-1}
\xrightarrow{\vec\psi \times}
\SQFT_{SU(2)}^{n,k}
\xrightarrow{\forget}
\SQFT^{n}
\xrightarrow{\nu(k)\times}
\SQFT_{SU(2)}^{n-3,k-1}
\xrightarrow{\vec\psi\times}\cdots
\quad.
\label{eq:seq}
\end{equation}
One of our main aims in this paper is to provide  a physics derivation of this long exact sequence,\footnote{%
The TMF version of this long exact sequence played an important role in \cite{LinYamashita},
but it is not that its derivation itself was one of the major results, since the corresponding long exact sequence 
exists for any genuinely equivariant generalized cohomology theories,
e.g. as the Gysin sequence for the spherical fibration $S^3\to *$ 
with the equivariance under $SU(2)$.
This means that, if we were able to show conclusively that the equivalence classes of SQFTs with symmetries form 
a genuinely equivariant generalized cohomology theory, 
the validity of the long exact sequence \eqref{eq:seq} would automatically follow, and 
there would be no independent need to
give a derivation as in this paper.
Here the intension is to give a derivation of this long exact sequence \eqref{eq:seq} directly within 
the current understanding of the groups $\SQFT^n$ and $\SQFT^{n,k}_{SU(2)}$.
}
using only the technique of supersymmetric quantum field theory
without utilizing the proposal \eqref{conj} of Stolz and Teichner.

We then use this sequence to study the following question.
Consider a heterotic compactification down to four dimensions, with $SU(2)$ symmetry with level one on the worldsheet.
We will have $SU(2)$ gauge symmetry in the spacetime.
Is the theory free of the Witten anomaly? 
This was argued to be the case in \cite{Tachikawa:2021mby,Yonekura:2022reu}, 
as part of a very general claim that no global anomalies are present in any heterotic compactifications in any dimensions.
But  no analysis specific to the case of the Witten anomaly was given in either.
The worldsheet theory in such a compactification gives rise to a class in $\SQFT^{26,1}_{SU(2)}$.
We explicitly compute this group, and 
find that massless Weyl fermions in each even-dimensional $SU(2)$ irreducible representation
always appear in pairs, from which we easily conclude that no Witten anomaly arises.
Along the way, we will need to invoke the conjecture \eqref{conj} a number of times,
but we will make explicit when this is done.

\paragraph{Organization of the paper.}
The rest of the paper is organized as follows.
In Sec.~\ref{sec:pre}, we begin by specifying the equivalence relation we use,
and introducing the groups $\SQFT^n$ and $\SQFT^{n,k}_{SU(2)}$.
We then discuss a couple of equivalence classes,
and derive a couple of useful equivalence relations, both of which will be needed later.

We then provide  a physics derivation of the long exact sequence \eqref{eq:seq} in Sec.~\ref{sec:seq}.
Denoting the homomorphisms connecting the Abelian groups in \eqref{eq:seq} uniformly by $d$,
what we need to do is first to show that $d^2=0$ and then to establish that $\Ker d=\Im d$.
We will discuss these issues in turn.

Finally, in Sec.~\ref{sec:app}, we discuss some applications of the long exact sequence \eqref{eq:seq}.
First, we will show that $\SQFT^{n,-1}_{SU(2)}\simeq 0$, using \eqref{eq:seq}.
Then, we will explain how $\SQFT^{n,1}_{SU(2)}$ can be computed from the knowledge of $\SQFT^m$, again using \eqref{eq:seq}.
Finally, we restrict our attention to the case $n=26$, which is relevant 
for the heterotic compactifications to four dimensions.
We will determine $\SQFT^{26,1}_{SU(2)}$, using the conjectural relation \eqref{conj} a couple of times.
We will then translate this information into the number of massless Weyl fermions in each irreducible representation of $SU(2)$, 
and find that each even-dimensional irreducible representation of $SU(2)$ appears even times, implying the absence of the Witten anomaly.

\paragraph{Remarks.}
Before proceeding, two remarks are in order.
First, this paper does not even cover the tip of the iceberg that is \cite{LinYamashita},
from which much more information on SQFTs can definitely be extracted.
The author would like to urge the readers of this paper to do so.\footnote{%
Y.-H. Lin and M. Yamashita kindly informed the author that a sequel \cite{LinYamashita2} to \cite{LinYamashita} is in preparation, in which physics translations 
(but not necessarily physics derivations) of the contents of  \cite{LinYamashita} 
will be provided.
The information presented there will include the general identification of $\nu(k)$
appearing in \eqref{eq:seq} and the analogous classes  for other long exact sequences
as \Nequals{(0,1)} ungauged and gauged Wess-Zumino-Witten models.
This information would be useful for anyone trying to generalize the physics derivation
of \eqref{eq:seq} in this paper to other long exact sequences in \cite{LinYamashita}.
}

Second,
in the opinion of the author, the long exact sequence \eqref{eq:seq} is noteworthy also from a historical point
of view.
In the past, physicists were quite content in studying each quantum field theory in isolation.
Gradually, they got used to considering families of quantum field theories, 
as in the case of the study of renormalization group flows,
and equivalence classes of quantum field theories, as in the study of symmetry protected topological phases.
Physicists are now accustomed to the study of  the structure of  operations on  quantum field theories,
such as the $SL(2,\bZ)$ transformations on the set of $U(1)$-symmetric theories in three dimensions \cite{Witten:2003ya}.
Long exact sequences formed by groups of equivalence classes of quantum field theories,
where the homomorphisms between groups are given by operations on quantum field theories,
can be considered as a natural next step.
That such sequences not only exist but can also be used to gain useful information
was a  pleasant surprise to the author.

\section{Preliminaries}
\label{sec:pre}

\subsection{The equivalence relation}
We start by reviewing the equivalence relation we introduce on the space of SQFTs.\footnote{%
The particular equivalence relation used here is chosen so that the results on the side of
topological modular forms can be reproduced on the side of SQFTs.
As such, details of the definitions given below might require 
further refinements and adjustments as the research progresses, 
although the author believes that 
the version presented here points toward the right direction.
}
Very roughly, we identify two SQFTs related by continuous deformations.
We also regard any SQFT which breaks supersymmetry spontaneously as `null'.
But we need a somewhat more sophisticated formulation than that for our purposes.
The concept has been
discussed in \cite{Gaiotto:2019asa,Yonekura:2022reu}
and in detail in \cite[Sec.3.1]{TachikawaYonekura},
so we will be brief.
This section is present mainly to set up our notations, and to make the paper 
a little more self-contained.

In this paper, we only consider unitary theories.
Then, the gravitational anomaly of a two-dimensional spin theory 
is parameterized by an integer $n$. 
We take the convention that $n=+1$ corresponds to a single Fermi multiplet,
where we  take the convention that  supersymmetry of \Nequals{(0,1)} theories are on the right-movers.

We also consider internal symmetries $G$, although we only consider the case $G=SU(2)$ in this paper.
The anomaly of a single $SU(2)$ is parameterized by  an integer $k$.
Our normalization is that a complex Fermi multiplet in the doublet of $SU(2)$ has $k=+1$.
In a conformal theory, this integer $k$ is the difference $k_L-k_R$
of the levels $k_{L,R}$ of left-moving and right-moving $SU(2)$ current algebras, respectively.
Unitarity restricts $k_{L,R}$ to be non-negative, but there is no such restriction on $k$.
As a slight abuse of the terminology, we also call this integer $k$
characterizing the anomaly of an $SU(2)$ symmetry the level of this symmetry,
and sometimes use the abbreviation $SU(2)_k$ to denote such symmetry.

\subsubsection{Compact and mildly noncompact theories}
As the zeroth step, we demand that the SQFTs we are going to classify 
are \emph{compact}, in the sense that the spectrum of their states is sufficiently discrete,
as in the case of sigma models on compact manifolds.
To define the equivalence relation  on two such compact SQFTs,
we need the concept of \emph{mildly non-compact} SQFTs.
Given a theory $S$ and a (possibly composite) chiral operator $\Phi$ of $S$,
let us introduce an additional Fermi multiplet $\Lambda_-$ and a parameter $\phi$, with
the superpotential interaction  \begin{equation}
m\int d\theta^+  \Lambda_-(\Phi-\phi),
\label{superp}
\end{equation}
where $m$ is a positive real parameter. 
Let us then take the limit $m\to +\infty$.
The path integral over $\Lambda_-$ forces the condition $\vev{\Phi}=\phi$ on the supersymmetric vacua.
As such, the resulting theory can be roughly considered as  `the theory $S$ 
with the vacuum expectation value of $\Phi$ fixed to be $\phi$'.
Let us denote the theory defined in this manner by $S_{\vev{\Phi}=\phi}$.

This theory $S$ is called \emph{mildly non-compact} 
if 
\begin{itemize}
\item $S_{\vev{\Phi}=\phi}$ is compact for all $\phi$,
\item $S_{\vev{\Phi}=\phi}$ becomes a theory $T_- $
independent of $\phi$ when $\phi < \phi_-$ for some $\phi_-$,
\item $S_{\vev{\Phi}=\phi}$ becomes a  theory $T_+$ 
independent of $\phi$ when $\phi > \phi_+$ for some $\phi_+$.
\end{itemize}
In such a case, $T_-$ and $T_+$ are called the \emph{incoming boundary}
and the \emph{outgoing boundary} of $S$, respectively.
Note that when $T_\pm$ has the gravitational anomaly $n$,
$S$ has the gravitational anomaly $n-1$.
Also, when we impose internal symmetries, $T_\pm$ and $S$ have the same level $k$.
A drawing of the theory $S$ together with associated objects 
can be given as follows:
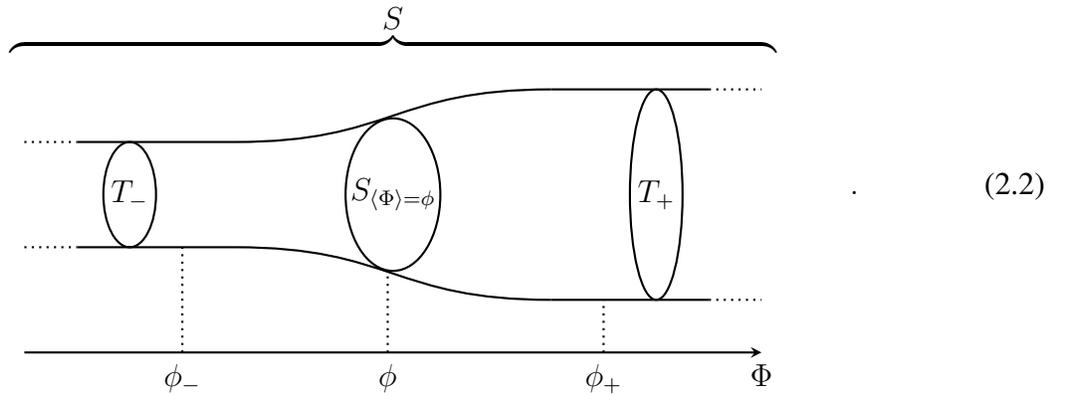
\begin{equation}
\begin{tikzpicture}[scale=.7,baseline=(A)]
    \node at (0,3) {$\overbrace{\phantom{aaaaaaaaaaaaaaaaaaaaaaaaaaaaaaaaaaaaaaaaaaaaaaa}}^{\displaystyle S}$};
    \draw[thick] (-5,0) ellipse (.5 and 1) node {$T_-$};
    \draw[thick,dotted] (-4,-3) node[below] {$\phi_-$} -- (-4,-1);
    \draw[thick] (0,0) ellipse (.9 and 1.45) node {$S_{\langle\Phi\rangle=\phi}$};
    \draw[thick,dotted] (-0.1,-3) node[below] {$\phi$} -- (-0.1,-1.55);
    \draw[thick,dotted] (+4,-3) node[below] {$\phi_+$} -- (+4,-2);
    \node (A) at (0,0) {};
    \draw[thick] (+5,0) ellipse (.5 and 2) node {$T_+$};
    \draw[thick,dotted] (-7,1) -- (-6,1);
    \draw[thick] (-6,1) -- (-3,1);
    \draw[thick,dotted] (-7,-1) -- (-6,-1);
    \draw[thick] (-6,-1) -- (-3,-1);
    \draw[thick] (3,2) -- (6,2);
    \draw[thick,dotted] (6,2) -- (7,2);
    \draw[thick] (3,-2) -- (6,-2);
    \draw[thick,dotted] (6,-2) -- (7,-2);
    \draw[thick,->] (-7,-3) -- (7,-3) node[below] {$\Phi$};
    \draw[thick] (-3,1) .. controls (0,1) and (0,2) .. (3,2);
    \draw[thick] (-3,-1) .. controls (0,-1) and (0,-2) .. (3,-2);
\end{tikzpicture}\qquad.
\end{equation}

Furthermore, we allow $T_+$  to be \emph{empty},
in the following sense.
Namely, the theory $S_{\vev{\Phi}=\phi}$ at $\phi>\phi_+$ 
breaks supersymmetry spontaneously
with the vacuum energy $E_0(\phi)>0$, and $E_0(\phi)\to +\infty$ as $\phi\to\infty$
sufficiently quickly. 
We regard a formal limit of $S_{\vev{\Phi=\phi}}$ as $\phi\to +\infty$ as an empty theory,
since no state exists below any given finite energy $E$ if we take $\phi$ to be large enough.
We allow $T_-$ to be empty in a similar manner.
We denote the empty theory symbolically by $0$.

\subsubsection{The equivalence relation}
In  analogy with the bordism relations of manifolds, then,
we declare two compact theories $T_1$ and $T_2$ to be equivalent in our sense
when there is a mildly non-compact theory $S$ whose incoming boundary is $T_1$
and whose outgoing boundary is $T_2$.
We summarize this situation by writing $S:T_1\to T_2$,
and we simply write $T_1\sim T_2$ if we are not particularly interested in $S$.
We also say that such an $S$ to provide a bordism between $T_1$ and $T_2$.

Below, we often say $T_1$ and $T_2$ are \emph{bordant} when $T_1\sim T_2$,
to emphasize that we are using our particular and peculiar equivalence relation.
When $T\sim 0$, we say that $T$ is \emph{null bordant}, or more simply, that it is \emph{null}.
Now, let us argue that this relation $\sim$ is actually an equivalence relation.

\paragraph{Reflexivity.}
It is clear that $T\sim T$. The necessary bordism is given by the theory $S$ obtained by adding a single 
chiral superfield $\Phi$ to $T$ without any interaction.

\paragraph{Transitivity.}
We assume that if  $T_1\sim T_2$ and $T_2\sim T_3$, we have $T_1\sim T_3$.
A rough idea is the following.
Suppose we are given 
a mildly noncompact theory $S$ with a chiral superfield $\Phi$
such that $S_{\vev{\Phi}\to-\infty} =T_1$ and $S_{\vev{\Phi}\to+\infty} =T_2$,
and another mildly noncompact theory $S'$ with a chiral superfield $\Phi'$
such that $S'_{\vev{\Phi'}\to-\infty} =T_2$ and $S'_{\vev{\Phi'}\to+\infty} =T_3$.
Then, it would be possible to `glue' the theories $S$ and $S'$ and form another theory $S'':=S\cup S'$,
by `gluing' two fields $\Phi$ and $\Phi'$ together
to form a single field $\Phi''$, so that \begin{equation}
S''_{\vev{\Phi''}\to-\infty}=T_1,\qquad
S''_{\vev{\Phi''}\sim 0}=T_2,\qquad
S''_{\vev{\Phi''}\to+\infty}=T_3.
\end{equation}

This sounds very plausible to the author, but a better argument will definitely be welcomed.
Note that this construction presupposes a kind of `locality' in the target space of a scalar field,
while the locality which is more often discussed is the locality in the spacetime.

\paragraph{Symmetry.}
Let us argue that $T_1\sim T_2$ implies $T_2\sim T_1$.
Suppose that there is a mildly noncompact theory $S$ such that $S_{\vev{\Phi}\to-\infty}=T_1$
and $S_{\vev{\Phi}\to+\infty}=T_2$.
We now define $\Phi':=-\Phi$. 
Then $S_{\vev{\Phi'}=-\phi}$ is obtained by adding to $S$ the superpotential \begin{equation}
m\int d\theta^+ \Lambda_- (-\Phi+\phi) \label{superp'}
\end{equation} instead. 
The fermion mass term in \eqref{superp'} has the   sign opposite to that of  \eqref{superp}.
Therefore, in the  limit $m\to +\infty$, there is a difference by the Arf theory,
the only nontrivial spin invertible theory in two dimensions.\footnote{%
For more on the Arf theory, see e.g.~\cite{Kaidi:2019tyf}.
}
Denoting a theory $T$ stacked with an Arf theory by $\overline{T}$ in general, we found that \begin{equation}
\overline{S}_{\vev{\Phi'}=-\phi} = S_{\vev{\Phi}=\phi},
\end{equation} meaning that \begin{equation}
\overline{S}_{\vev{\Phi'}\to -\infty}=T_2,\qquad
\overline{S}_{\vev{\Phi'}\to +\infty}=T_1.
\end{equation}
So we indeed found $T_2\sim T_1$.

\subsubsection{The graded-commutative rings of equivalence classes}


Given theories $T$ and $T'$, we denote by $T+T'$ and $T\cdot T'$
the theories whose Hilbert spaces are the direct sum and the tensor product of the two theories involved, respectively.
It is fairly clear that the equivalence relation $T_1\sim T_2$ introduced above
respects the addition   and the multiplication of theories. 
The addition and the multiplication then descend consistently to those on the equivalence classes $[T]$ under the relation $\sim$.

Now we argue that $0\sim T+\overline{T}$, where we remind the reader that $\overline{T}$ is the theory
$T$ multiplied by the Arf theory.
To show this, we simply consider adding a chiral superfield $\Xi$ to $T$ to define the theory $S$,
and use $\Phi:=\Xi^2$ as the coordinate to slice the combined theory.
The superpotential interaction is then \begin{equation}
m\int d\theta^+\Lambda_-( \Xi^2 - \phi).
\end{equation}
When $\phi\ll 0$, there is no supersymmetric vacua, and therefore $S_{\vev{\Phi}\to-\infty}=0$.
When $\phi\gg 0$, there are two supersymmetric vacua around $\vev{\Xi}\simeq \pm\sqrt{\phi}$.
The fermion mass term there is proportional to $\vev{\Xi}$,
and therefore has opposite signs at the two vacua,
leading to a multiplicative difference by the Arf theory between the two.
Therefore $S_{\vev{\Phi}\to+\infty}=T+\overline{T}$.
This means that $[\overline{T}]$ gives an additive inverse to $[T]$ at the level of the equivalence classes.
This in turn implies that the equivalence classes form an Abelian group.

As for the product, it is close to commutative but not quite, see \cite[Sec.~3.1]{TachikawaYonekura} for details.
This small subtlety makes the product on the equivalence classes graded-commutative.

In the discussions so far, we did not specify how much symmetry we impose on the theories involved.
When we demand all theories involved to have \Nequals{(0,1)} supersymmetry only,
we have the graded-commutative ring of equivalence classes, which we denote by $\SQFT$.
Similarly, when we demand all theories involved to have \Nequals{(0,1)} supersymmetry and an internal symmetry $G$,
we have the graded-commutative ring of equivalence classes, which we denote by $\SQFT_G$.
Many person-hours have been spent in the study of $\SQFT$.
The aim of this paper is to initiate the study of $\SQFT_G$.

We note that $\SQFT$ and $\SQFT_G$ are graded by the anomalies of the theories involved.
We then have Abelian groups $\SQFT^n$ and $\SQFT^{n,k}_{SU(2)}$,
obtained by restricting to theories with the specified gravitational and $SU(2)$ anomalies.
As the multiplication of theories adds anomalies, we have products \begin{align}
\SQFT^n \times \SQFT^{n'}&\to \SQFT^{n+n'},\\
\SQFT_{SU(2)}^{n,k} \times \SQFT_{SU(2)}^{n',k'}&\to \SQFT_{SU(2)}^{n+n',k+k'}.
\end{align}
We also have an obvious homomorphism \begin{equation}
\SQFT_{SU(2)}^{n,k}\xrightarrow{\forget} \SQFT^n
\end{equation} given by forgetting the $SU(2)$ symmetry.


\subsection{Properties of the equivalence relation}
\label{subsec:prop}

Let us now discuss a few properties which we will need to use later.

\paragraph{Geometric bordism gives SQFT bordism.}
One way to generate a large class of SQFT equivalences is to use
mildly non-compact sigma models.
Take  a $(d+1)$-dimensional manifold $N_{d+1}$ 
with an incoming boundary $M_d$ and an outgoing boundary $M_{d}'$.
Let us say that it is equipped with a metric and an appropriate $B$ field,
so that there is no sigma model anomaly.
Then, the \Nequals{(0,1)} sigma model $\sigma(N_{d+1})$ on $N_{d+1}$
has an incoming boundary theory $\sigma(M_d)$ and 
an outgoing boundary theory $\sigma(M_d')$,
giving rise to an SQFT bordism $\sigma(M_d)\sim \sigma(M_d')$.

A small variant of this construction is the following. 
We  further equip the manifold $N_{d+1}$ with a $G$-bundle with connection $A$.
We take another theory $T$  with $G$ symmetry, and try to fiber the theory $T$ 
using this $G$-connection $A$.
The cancellation of the sigma model anomaly requires us that a certain topological condition
is met, so that we can pick an appropriate $B$ field satisfying a number of consistency conditions.\footnote{%
One of such consistency conditions when $G=SU(2)$ is the requirement 
$dH=p_1(R)/2 + kc_2(G)$ at the differential form level, 
where $p_1(R)$ is the Pontryagin class of the curvature of the sigma model target space 
and $c_2(G)$ is the second Chern class of the $G$-bundle.
A more precise set of conditions was first studied in \cite{Witten:1985mj}
and was carefully explained in \cite{Yonekura:2022reu}.
Mathematically, this set of data is known as a (differential) twisted string structure.}
Once such a $B$-field is chosen,
we have a mildly-compact SQFT which we can denote by $\sigma(N_{d+1}\times_G T)$,
providing an SQFT bordism $\sigma(M_d \times_G T)\sim \sigma(M_d'\times_G T)$.


\paragraph{Superpotential deformation as an SQFT bordism.}
Suppose that the theory  $T$ has a (possibly composite) Fermi multiplet $X_-$.
Let $T'$ be the theory obtained by adding to the theory $T$
the superpotential term 
\begin{equation}
\int d\theta^+ X_-. \label{superp-}
\end{equation}
Then we have $T\sim T'$.
To see this,  we add a chiral superfield $\Phi$ to the theory $T$, 
and add the superpotential term 
\begin{equation}
\int d\theta^+ f(\Phi)X_-,
\end{equation}
where $f(\phi)$ is a function such that $f(\phi)\to 0$ when $\phi\to -\infty$ and $f(\phi)\to 1$ when $\phi\to +\infty$.
Call the resulting theory $S$. Then $S_{\vev{\Phi}\to-\infty}$ is the original theory 
$T$ and $S_{\vev{\Phi}\to+\infty}$ is 
the theory $T'$ obtained by adding the superpotential term \eqref{superp-}.

\paragraph{Spontaneous supersymmetry breaking and the null bordism.}

Next, suppose that a theory $T$ breaks supersymmetry spontaneously. 
Then we would like to show that $T\sim 0$,
i.e.~there is a mildly noncompact theory $S$ such that
$S_{\vev{\Phi}\to-\infty}=T$ 
and $S_{\vev{\Phi}\to+\infty}$ is empty.
The idea is the following. 

Let $V_0>0$ be the vacuum energy density of the theory $T$.
Now, take a dimensionless parameter $c$ and scale the theory by this factor,
so that the vacuum energy density is $cV_0 >0$.
Denote this rescaled theory by $T(c)$.
When $c$ is close to one, $c=1+\delta c$,
this rescaling can be done by adding to the Lagrangian the trace $T_{\mu\mu}$
of the energy momentum tensor,
which is known to have a supersymmetric version \begin{equation}
(\delta c )\int d\theta^+  J_-,
\end{equation} where $J_-$ is the supercurrent for the super-Weyl transformation.\footnote{%
For a recent summary of supercurrents of two-dimensional \Nequals{(0,1)} systems,
see e.g.~\cite{Baggio:2018rpv}.}

Now we introduce a function $f(\phi)$ 
such that $f(\phi)\to 1$ when $\phi\to -\infty$
 and $f(\phi)\to +\infty$ sufficiently quickly when $\phi\to +\infty$.
We consider the parameterized theory $T(f(\phi))$, 
so that it approaches the original theory $T$ when $\phi\to -\infty$
and becomes empty when $\phi\to+\infty$.
We then promote this parameter $\phi$ to a dynamical chiral superfield $\Phi$, and obtain the theory $S$.
Very schematically, this corresponds to the addition to the Lagrangian of the interaction \begin{equation}
\int d\theta^+ (f(\Phi) -1 )J_-.
\end{equation}
This sounds very plausible to the author, but again, a better derivation will be necessary and welcomed.

\paragraph{Renormalization group flow as an SQFT bordism.}
The discussion above can be generalized to the case when $T$ does not necessarily break supersymmetry.
In that case, the same construction gives a mildly-noncompact theory such that $S_{\vev{\Phi}\to -\infty}=T$
and $S_{\vev{\Phi}\to +\infty}$ is the low-energy limit of $T$.
Therefore, any theory $T$ is bordant to its low-energy limit, in our definition.

\paragraph{Compact theories from two null bordisms.}
Next, we suppose we found two null bordisms $S: T\to 0$ and $S':T\to 0$ for a single theory $T$.
As we discussed, $\overline{S'}$ gives us a bordism in the opposite direction, $\overline{S'}: 0\to T$.
Then the glued theory  $\overline{S'}\cup S$ gives a bordism $0\to  0$.
We assume that this gives rise to a compact theory,
as both regions of $\vev{\Phi}\to \pm\infty$ have arbitrarily large vacuum energy,
effectively making the $\Phi$ direction compact.

We will repeatedly use this construction below.
Note that when the theory $T$ has gravitational anomaly $n$,
the theory $\overline{S'}\cup S$ has gravitational anomaly $n-1$.
The level $k$ of the internal symmetry $G$ is unchanged.

\subsection{Some concrete equivalence classes}
\label{sec:concrete}

We now introduce a couple of concrete equivalence classes which will be needed later.

\paragraph{Free Fermi multiplets.}
We start with a theory of four real Fermi multiplets. 
It has $SO(4)$ symmetry rotating four fermions.
Therefore it  has $SU(2)\times SU(2)$ symmetry, and the levels are both one.
We denote the resulting bordism class by 
\begin{equation}
\vec\psi\in \SQFT^{-4,1,1}_{SU(2)\times SU(2)}.
\end{equation}
In this paper we only use one of the two $SU(2)$ symmetries, so we regard it as an element \begin{equation}
\vec\psi\in \SQFT^{-4,1}_{SU(2)}.
\end{equation}

Below, we will need to describe superpotential deformations in SQFTs
which has this theory $\vec\psi$ as one of the ingredients. 
In these cases, we will use the symbol $\vec\Psi_-$ for the four real Fermi multiplets contained in it.

\paragraph{Sigma models on $S^3$.}
The next theory we need is the \Nequals{(0,1)} sigma model on a round $S^3$ with the $B$-field
whose field strength $H$ has the flux $\int_{S^3} H=k$.
This has a geometric action of $SU(2)\times SU(2)$ given by the left multiplication and the right multiplication of $S^3\simeq SU(2)$.
As was nicely reviewed in \cite[Sec.~2]{Gaiotto:2019asa}, 
it gives a theory with $SU(2)_L\times SU(2)_R$ symmetry with levels $k-1$ and $-k-1$, respectively,
determining a bordism class we denote by \begin{equation}
\nu(k) \in \SQFT^{3,k-1,-k-1}_{SU(2)\times SU(2)}.
\end{equation}
The low energy limit is believed to behave as follows:
\begin{itemize}
\item When $|k|\ge 1$, 
it is a product of bosonic diagonal $SU(2)_{|k|-1}$ conformal field theory (CFT) together with three right-moving fermions in the adjoint of $SU(2)$. 
This is supersymmetric, with an $SU(2)_\ell$ symmetry on the left-movers with level $|k|-1$
and another $SU(2)_r$ symmetry on the right-movers with level $-(|k|-1+2)=-|k|-1$,
where $+2$ is from the adjoint fermion.
Note that the identification of the sigma model symmetries $SU(2)_L\times SU(2)_R$
and the CFT symmetries $SU(2)_\ell\times SU(2)_r$ is such that \begin{align}
SU(2)_L &\simeq SU(2)_\ell, &
SU(2)_R &\simeq SU(2)_r, & (k&\ge 1), \\
SU(2)_L &\simeq SU(2)_r, &
SU(2)_R &\simeq SU(2)_\ell, & (k&\le -1).
\end{align}
\item In particular, when $k=\pm 1$, the infrared theory consists solely of three right-moving free fermions $\psi_{a=1,2,3}$, with the supercharge given by $Q\propto \psi_1\psi_2\psi_3$.
In our convention, this theory has the $SU(2)$ level $-2$.
\item 
When $k=0$, this is simply the $S^3$  sigma model without the $B$-field. 
This can be realized in terms of four chiral superfields $\Phi_{i=1,\ldots,4}$  and a Fermi multiplet $\Lambda_-$,
with the superpotential \begin{equation}
c\int d\theta^+ \Lambda_-(|\Phi|^2-c') 
\end{equation}  with $c$, $c'$ some constants.
The renormalization group drives $c'$ to a large negative value, breaking supersymmetry.
The low energy limit will then simply be a left-moving goldstino
and four real Majorana right-moving fermions, producing two $SU(2)$ level $-1$ symmetries.
So we have found \begin{equation}
\nu(0)\sim 0.
\end{equation}
\end{itemize}
In this paper we only use one of the two $SU(2)$ symmetries of $\nu(k)$,
and regard it as a class in $\SQFT^{-3,k-1}_{SU(2)}$.

\subsection{Two lemmas}
\label{sec:lemmas}

Next we discuss two lemmas which will be used later.
They also serve as illustrations of the constructions related to the equivalence relation
we have discussed so far.

\subsubsection{First lemma}
\label{sec:first-lemma}

We start by considering the geometry $S^3\times \bR$.
Parameterize the $\bR$ direction by $\phi$.
We equip it with an $SU(2)$-bundle with $\int c_2=1$ concentrated around $\phi\sim 0$,
and put the $B$-field satisfying $dH=\ell c_2$.
Now, given a theory $T$ with $SU(2)$ level $\ell$,
we can fiber it over the sigma model on $S^3\times \bR$ with this $SU(2)$ gauge field and the $B$-field.
Let us call the total mildly-noncompact theory by $S$.
This is a slight generalization of the standard construction of $SU(2)$ 
gauge configurations in heterotic string theory,
where $\ell=1$ and $T$ is the left-moving current algebra theory with central charge $c_L=16$.

We  have $\int_{S^3} H_{\phi\to+\infty} -\int_{S^3}H_{\phi\to-\infty}=\ell$.
Let us then write \begin{equation}
k-\ell=\int_{S^3} H_{\phi\to-\infty},\qquad
k=\int_{S^3} H_{\phi\to+\infty}.
\end{equation}
On the boundary, the theory is a simple product $\nu(k-\ell) T$ or $\nu(k) T$.
Then $S$ provides a bordism \begin{equation}
S\colon \nu(k-\ell)T\to \nu(k)T.
\end{equation}

We can actually make this bordism $S$ preserve two $SU(2)$ symmetries.
To see this,
we start from the trivial principal bundles $(S^3\times \bR_{<0})\times SU(2)$
and $(S^3\times \bR_{>0})\times SU(2)$,
where $S^3$ is part of the base and $SU(2)$ is the fiber.
We paste them at $\phi=0$,
using  the identify map $S^3\to SU(2)$ as the gauge transformation.
Then, the point  $(x,g_-)\in S^3\times SU(2)$ at $\phi<0$
is mapped to the point $(x,g_+)\in S^3\times SU(2)$ at $\phi>0$ via 
\begin{equation}
g_-=x g_+. \label{mapping}
\end{equation}
The action of $(g,g')\in SU(2)^2$ given by \begin{equation}
(x,g_-)\mapsto ( gx(g')^{-1} , g g_-   )
\label{action1}
\end{equation} is mapped to \begin{equation}
(x,g_+)\mapsto ( gx(g')^{-1} ,g'  g_+ ).
\label{action2}
\end{equation}
So far we have only specified the bundle structure,
but we can further equip it with a connection whose curvature is 
concentrated around $\phi\sim 0$, and an $H$-flux which is spherically symmetric.

Note that $\nu(k-\ell)T$ has three $SU(2)$ symmetries with levels $k-\ell-1$, $-(k-\ell)-1$ from $\nu(k-\ell)$
and $\ell$ from $T$.
Call them $SU(2)_L$, $SU(2)_R$ and $SU(2)_T$.
Similarly $\nu(k)T$ has three $SU(2)$ symmetries with level $k-1$, $-k-1$ and $\ell$.
Call them $SU(2)_L'$, $SU(2)_R'$ and $SU(2)_T$.
Comparing \eqref{action1} and \eqref{action2}, we find that
the two $SU(2)$ symmetries preserved by the bordism $S$ above are
  \begin{equation}
\text{diag}(SU(2)_{L}\times SU(2)_T) \simeq SU(2)_{L}', 
\label{ident1}
\end{equation} whose levels are $(k-\ell-1)+\ell=k-1$,
and \begin{equation}
SU(2)_{R} \simeq \text{diag}(SU(2)'_{R}\times SU(2)_T)
\label{ident2}
\end{equation} whose levels are $-(k-\ell)-1=-k-1+\ell$.
We summarize this by writing \begin{equation}
_{\color{blue} k-\ell-1}\nu(k-\ell)_{\color{red}-k+\ell-1} T_{\color{blue} \ell} 
\quad \sim \quad
_{\color{blue} k-1}\nu(k)_{\color{red} -k-1} T_{\color{red} \ell},
\label{main-trick}
\end{equation}
where levels of various $SU(2)$ symmetries are given as subscripts,
with the colors distinguishing which of \eqref{ident1} and \eqref{ident2} they contribute to.

As a figure, the bordism can be summarized in the following manner:
\begin{equation}
\vcenter{\hbox{\begin{tikzpicture}[scale=.7]
\shade[left color=white,right color=white,middle color=pink] (-4,1) rectangle (-2, -1) ;

\draw[thick] (0,0) ellipse (1 and 1) node {$k$};
\draw[thick] (-6,0) ellipse (1 and 1) node {$k-\ell$};
\draw[thick] (-6,1) -- (0,1);
\draw[thick] (-6,-1) -- (0,-1);
\node (A) at (-3,2.5)  {$\displaystyle\int c_2=1$};
\draw[->] (A) -- (-3,0);
\node at (0,2.5)  {$T$};
\node at (0,1.5)  {$\times$};
\node[red] at (3,2.5)  {$\ell$};
\node[red] at (3,1)  {$-k-1$};
\node[blue] at (3,-1)  {$+k-1$};
\node at (-6,2.5)  {$T$};
\node at (-6,1.5)  {$\times$};
\node[blue] at (-9,2.5)  {$\ell$};
\node[red] at (-9,1)  {$-k+\ell-1$};
\node[blue] at (-9,-1)  {$+k-\ell-1$};
\end{tikzpicture}}}\quad .
\end{equation}
Here,  a circle with a number $x$ is for $S^3$ with $\int_{S^3} H=x$,
and a shaded region in the middle shows an $SU(2)$ background with $\int c_2=1$.
We also indicated the levels of various $SU(2)$ symmetries on the boundary theories
in it.
From the top to the bottom, the levels are for $SU(2)_T$, $SU(2)_{R}$, $SU(2)_{L}$ on the left,
and for $SU(2)_T$, $SU(2)_{R}'$, $SU(2)_{L}'$ on the right.
We also used different colors to specify which of the preserved $SU(2)$ symmetries 
\eqref{ident1} and \eqref{ident2} they contribute to.

\subsubsection{Second lemma}
\label{second-lemma}

Next, we show that  the forgetful map $ \forget: \SQFT^{n,0}_{SU(2)} \to \SQFT^n$
is actually an isomorphism, so that we actually have \begin{equation}
\forget: \SQFT^{n,0}_{SU(2)} \xrightarrow{\sim}\SQFT^n.
\end{equation}
To show this, we need to construct an inverse, for which we use the map
\begin{equation}
R:\SQFT^n\to \SQFT^{n,0}_{SU(2)}
\end{equation}
 simply regarding a theory $T$ to have a trivially-acting $SU(2)$ symmetry.
It is obvious that $ 
T=\forget(R(T)).
$ 
So we need to establish that 
$ 
T\sim R(\forget(T))
$ 
for $T\in \SQFT^{n,0}_{SU(2)}$.
In other words, if we denote by $T[A]$ the theory $T$ coupled to the $SU(2)$ background gauge field $A$, we would like to show that $ 
T[A] \sim T[0].
$ 

To argue for this, let us recall that the class $\nu(1)\in \SQFT^{-3,0,-2}_{SU(2)\times SU(2)}$ 
has two realizations, one as the $S^3$ sigma model with $\int_{S^3}H=1$,
another as three right-moving free fermions in the adjoint of $SU(2)$ of level $-2$.
Denote by $\nu(1)[A,A']$ the class $\nu(1)$ coupled to the background gauge fields
$A$ and $A'$ for the $SU(2)_0$  and $SU(2)_{-2}$ symmetries, respectively.
Using the free fermion realizations, it is clear that \begin{equation}
\nu(1)[A,A'] \sim \nu(1)[0,A'],
\end{equation} but this is not obvious in the sigma model description.

We now multiply both sides by $T[A']$ and gauge $A'$.
We then have 
\begin{equation}
\frac{\nu(1)[A,a']\times T[a']}{\text{gauge $a'$}} \sim
\frac{\nu(1)[0,a'] \times T[a']}{\text{gauge $a'$}} \label{nuTa}
\end{equation}
This gauging of the second $SU(2)$ is possible since the gauginos in the vector multiplet carry the level $+2$, 
so that the total level is $-2+2=0$,
meaning that this $SU(2)$ symmetry is non-anomalous.

Let us focus on the left hand side, as the right hand side can be obtained by setting $A=0$ at the end.
We want to analyze the theory \begin{equation}
\frac{\nu(1)[A,a']\times T[a']}{\text{gauge $a'$}}.
\end{equation}
Using the sigma model description, the gauge group can be completely fixed 
by setting the $S^3$ coordinate at some fixed value, say $e\in S^3$.
This sets the vacuum expectation value of $a'$ to be equal to $A$.
Fluctuations of $a'$ around $A$
combine with the fluctuations of $S^3$ around $e\in S^3$ and become massive.
In the very massive limit, we end up with just having $T[A]$.
Applying this construction to both sides of \eqref{nuTa}, we conclude \begin{equation}
T[A]\sim T[0],
\end{equation}
which was what we wanted to show.
Note that the same argument applies to other non-Abelian simple simply-connected Lie groups $G$, but not to $U(1)$.

\section{The long exact sequence}
\label{sec:seq}

After these preparations, we now move on to the derivation of the long exact sequence of Abelian groups
\begin{equation}
\cdots\xrightarrow{\nu(k)\times} 
\SQFT_{SU(2)}^{n-4,k-1}
\xrightarrow{\vec\psi \times}
\SQFT_{SU(2)}^{n,k}
\xrightarrow{\forget}
\SQFT^{n}
\xrightarrow{\nu(k)\times}
\SQFT_{SU(2)}^{n-3,k-1}
\xrightarrow{\vec\psi\times}\cdots .
\label{eq:seqx}
\end{equation}
If we denote all the operations uniformly by $d$,
what we need to establish is firstly $d^2=0$
and  then secondly $\Ker d=\Im d$.

\subsection{$d^2=0$}

\subsubsection{$\forget\circ (\vec\psi\times)$ is null}
\label{forget-psi}
This is because the theory  $\vec\psi$ of four left-moving Majorana-Weyl fermions is  null when $SU(2)$ action is forgotten.
To see this, simply  add the superpotential 
\begin{equation}
\propto \int d\theta^+ \vec c\cdot \vec\Psi_-
\end{equation}
 for some nonzero vector $\vec c$,
where $\vec\Psi_-$ stands for the four Fermi multiplets contained in this theory $\vec\psi$.
This provides a nonzero energy $\propto |\vec c|^2$, and breaks supersymmetry spontaneously.
The goldstino associated to this spontaneous breaking is 
the component of the Majorana-Weyl fermion parallel to $\vec c$.
Note that this deformation necessarily breaks the $SU(2)$ symmetry,
so this operation can be done only after forgetting the $SU(2)$ symmetry.

Recalling the discussion in Sec.~\ref{subsec:prop}, we should be able to represent this 
superpotential deformation as a null bordism  given by a mildly non-compact theory. 
Concretely, the null bordism can be constructed by adding a chiral superfield $\Phi$
and introducing the superpotential \begin{equation}
\propto \int d\theta^+ f(\Phi)\vec c\cdot \vec\Psi_-
\end{equation} 
for a suitable function $f(\phi)$ such that $f(\phi)\to 0$ when $\phi\to -\infty$ 
while $f(\phi)\to +\infty$ when $\phi\to +\infty$.

\subsubsection{$(\nu(k)\times ) \circ \forget$ is null}
\label{nu-forget}
To show this, we need to find a null bordism of $\nu(k)$ times a theory $T \in \SQFT_{SU(2)}^{n,k}$,
in a way keeping the $SU(2)_{k-1}$ symmetry of $\nu(k)$ intact.
For this purpose, we simply use the bordism \eqref{main-trick}  with $\ell=k$,
which was
\begin{equation}
\nu(0)T\sim \nu(k)T.
\end{equation}
This is what we need, because $\nu(0)\sim 0$.

In this case the bordism \eqref{main-trick} can also be  considered as a ball $B^4$ with a cigar-like metric,
with no incoming boundary and an outgoing $S^3$ boundary.
We have an $SU(2)$-instanton configuration within $B^4$ 
with $\int_{B^4} c_2=1$,
and the $B$-field solves $dH=kc_2$.
Therefore the $H$-flux on the outgoing $S^3$ boundary satisfies $\int_{S^3} H=k$.
This simplified setup can be visualized as follows:
\begin{equation}
\vcenter{\hbox{\begin{tikzpicture}[scale=.7]
\clip (-6,3.5) rectangle (1.2,-1.2);
\begin{scope}
\clip (0,0) ellipse (5 and 1);
\shade[inner color=pink] (-4.5,0) ellipse (1 and 1) node {};
\end{scope}
\draw[thick] (0,1) arc[start angle=90, end angle=270, x radius=5, y radius= 1];

\draw[thick] (0,0) ellipse (1 and 1) node {$k$};
\node (A) at (-4.5,2.5)  {$\displaystyle\int c_2=1$};
\draw[->] (A) -- (-4.5,0);
\node at (0,2.5)  {$T$};
\node at (0,1.5)  {$\times$};
\end{tikzpicture}}} \quad .
\label{kt-twisted}
\end{equation}
Note that the case where $k=1$ and $T$ is the $c_L=16$ current algebra theory
is the standard worldsheet description of the smooth NS5-brane in heterotic string theory,
realized as a one-instanton configuration.

\subsubsection{$(\vec\psi \times)\circ (\nu(k) \times)$ is null}
\label{psi-nu}
For this purpose, 
it suffices to show that  $\vec\psi \nu(k)$ is null as an element in $\SQFT_{SU(2)}^{1,k}$,
where the $SU(2)$ level $k$ is from the $SU(2)$ symmetry of $\vec\psi$ with level $1$ and 
$SU(2)_L$ symmetry of  $\nu(k)$ with level $k-1$.
Now,  note that the embedding of the unit sphere \begin{equation}
\vec X\colon S^3\to \bR^4
\end{equation} 
is $SU(2)_L$-equivariant.
Therefore,  the superpotential term\begin{equation}
\int d\theta^+ \vec X\cdot \vec \Psi_-
\end{equation} of the combined theory is $SU(2)$ invariant.
This breaks supersymmetry everywhere on $S^3$ since $\vec X$ is everywhere nonzero on $S^3$.
Therefore this theory is null.

\subsection{$\Ker d=\Im d$}

Now that we have shown $d^2=0$, let us show $\Ker d=\Im d$.
The general strategy is as follows.
Suppose we have a sequence of operations 
\begin{equation}
\mathrm{R'}\xrightarrow{C} \mathrm{P} \xrightarrow{A}\mathrm{Q}\xrightarrow{B} \mathrm{R} \xrightarrow{C} \mathrm{P}',
\label{PQR}
\end{equation}
where we already showed that $A\circ C$, $B \circ A$, $C\circ B$
are all null in general.
Further suppose $T\in \mathrm{Q}$ satisfies $B(T)\sim 0$.
In this case, we construct two null bordisms of $C\circ B(T)$:
\begin{itemize}
\item one $S: C\circ B(T)\sim 0$ uses the general null bordism of $C\circ B$, and
\item  another $S': C\circ B(T)\sim 0$ uses the assumption $B(T)\sim 0$.
\end{itemize}
Pasting them, we get a bordism $\overline{S} \cup S' : 0\sim 0$,
which gives rise to a compact theory in $\mathrm{P}$ which we denote by $T':=\langle C,B,T \rangle$.
We need to show $A(T')=A(\langle C,B,T \rangle)\sim T$,
but this can be done by using the general null bordism of $A\circ C$.\footnote{%
The rough explanation is as follows. 
The null bordism for $A\circ C$ can be readily applied on the side of $S'$,
since it only uses the null bordism $B(T)\sim 0$ which does not interfere 
with the null bordism for $A\circ C$.
But on the side of $S$, the two null bordisms, one for $C\circ B$ and another for $A\circ C$,
interfere with each other, and fail to disappear at one point, 
leaving only the original theory $T$ at the end.
What was discussed in this footnote would be more easily understandable
after the reader would have gone through the explicit constructions given below.
}
Now, let us discuss the three cases in turn.

\subsubsection{$T\sim \forget(T')$ when $\nu(k)T\sim 0$}
Suppose that $T\in \SQFT^n$  is such that $\nu(k)T\sim 0$. 
We create the theory $T'$ by pasting two null bordisms of $\vec\psi \nu(k) T$  as follows:
\begin{equation}
T':=\qquad \begin{tikzpicture}[scale=.7,baseline=(A).baseline]

\draw[thick] (1,-1) arc[start angle=-90, end angle=90, x radius=3, y radius= 1.8];

\draw[thick] (-0.7,1) arc[start angle=90, end angle=270, x radius=3, y radius= 1.9];

\draw[thick] (0,0) ellipse (1 and 1) node (A) {$k$};
\draw[->] (-4,-3) -- (+4,-3) node[below] {$\Phi$};
\node at (-1.5,-3.5) {$\underbrace{\phantom{aaaaaaaaa}}_{\displaystyle\overline{S}}$};
\node at (1.8,-3.5) {$\underbrace{\phantom{aaaaaaaaa}}_{\displaystyle S'}$};
\node at (0,2.5)  {$T$};
\node at (0,1.5)  {$\times$};
\node at (3,-1.5)  {$\times$};
\node at (3,-2.5)  {$\vec\psi$};
\node at (0,-1.5)  {$\times$};
\node at (0,-2.5)  {$\vec\psi$};
\end{tikzpicture}.
\end{equation}
On the right hand side, i.e.~on the side of $S'$, we use the null bordism  $\nu(k) T \sim 0$ which is part of the assumption.
On the left hand side, i.e.~on the side of $\overline{S}$, we use the general null bordism $\vec\psi \nu(k)\sim 0$ constructed in Sec.~\ref{psi-nu}.
Both have the same boundary $\vec\psi \nu(k) T$, so we can paste them to construct a compact SQFT $T'$.
By construction  $T'\in \SQFT_{SU(2)}^{n,k}$.
We also showed the additional direction $\bR_\Phi$ parameterized by $\Phi$,
the operator used to slice the mildly non-compact theories $\overline{S}$ and $S'$,  in the diagram above.
Then we have a superpotential \begin{equation}
\int d\theta^+ f(\Phi) (\vec X \cdot \vec \Psi_-) ,
\label{3.7}
\end{equation} where $f(\phi)$ is a monotonic function such that $f(\phi)\to \infty$ when $\phi\to -\infty$
and $f(\phi)=0$ when $\phi>0$.

We now need to check that this element gives back $T$ when we forget the $SU(2)$ action.
For this, we take the inspiration from the null bordism $\forget(\vec\psi)\sim 0$ we discussed in Sec.~\ref{forget-psi}.
Namely, we pick a nonzero constant $\vec X_0$ and add 
\begin{equation}
\int d\theta^+\vec X_0\cdot \vec\Psi_-
\label{3.8}
\end{equation} to the superpotential.
The total superpotential, the sum of \eqref{3.7} and \eqref{3.8}, 
provides  positive vacuum energy 
almost everywhere, except on the points where 
\begin{equation}
f(\Phi)\vec X +\vec X_0=0.
\label{va}
\end{equation}
This happens at a single point $p$ on $S^3\times \bR_\Phi$,
where $f(\Phi)=|\vec X_0|$ and $\vec X=-\vec X_0/|\vec X_0|$.
Parameterizing the   $\Phi$ direction and the  $S^3$ direction 
around the point $p$ uniformly by $\vec Y$
such that the point $p$ is at $\vec Y=0$, we essentially have a superpotential \begin{equation}
\int d\theta^+ \vec Y\cdot \vec\Psi_-.
\end{equation}
This removes the pair of  $\vec Y$ and $\vec \Psi_-$ by giving them a nonzero mass,
and only the theory $T$ remains in the infrared limit.

\subsubsection{$T\sim \nu(k)T'$ when $\vec\psi T\sim 0$}
Next, suppose that an element $T\in \SQFT_{SU(2)}^{n-4,k-1}$ is null after tensoring by $\vec\psi$.
In this case, we consider the SQFT  $T'(\vec C)\in \SQFT^{n-1}$ 
depending on a nonzero vector $\vec C\in \bR^4\simeq \bC^2$
given by 
\begin{equation}
T'(\vec C):=\quad \begin{tikzpicture}[scale=.7,baseline=(A)]
\draw[thick] (-0.2,1) arc[start angle=90, end angle=270, x radius=3, y radius= .5];
\node at (0,2.5)  {$T$};
\node (A) at (0,1.5)  {$\times$};
\node at (0,0.5)  {$\vec\psi$};

\draw[thick] (.5,0) arc[start angle=-90, end angle=90, x radius=3, y radius= 1.5];

\draw[->] (-4,-.7) -- (+4,-.7) node[below] {$\Phi$};
\node at (-1.5,-1.2) {$\underbrace{\phantom{aaaaaaaaa}}_{\displaystyle\overline{S}}$};
\node at (1.8,-1.2) {$\underbrace{\phantom{aaaaaaaaa}}_{\displaystyle S'}$};

\end{tikzpicture}\quad 
\label{would-be-compact}
\end{equation}
where we denoted the chiral superfield parameterizing the bordism direction again by $\Phi$.
Here, the bordism $S': \vec\psi T\sim 0$ on the right hand side is given as part of the assumption,
and the bordism $S: \vec\psi T\sim 0$ on the left hand side is what was discussed in Sec.~\ref{forget-psi}.
That is, we introduce the superpotential term \begin{equation}
\int d\theta^+  (\vec C \cdot \vec \Psi_-) f(\Phi),
\end{equation}
where  
$f(\phi)$ is such that $f(\phi)\to \infty$ as $\phi\to -\infty$
and $f(\phi)=0$ when $\phi>0$.
Note that the theory $T'(\vec C)$ can be defined for $\vec C=0$, but then the theory is mildly noncompact
with an incoming boundary at $\vev{\Phi}\to -\infty$.

We now multiply $T'(\vec C)$ by $\nu(k)$, and would like to show that it is bordant to the original theory $T$.
For this, we take the inspiration from the null bordism $(\nu(k)\times)\circ \forget\sim 0$ we discussed in Sec.~\ref{nu-forget},
which is  the  bordism \eqref{kt-twisted}. 
So, we take a ball $B^4$ with $\int c_2=1$ as before,
and try to fiber  the  compact theory $T'(\vec C)$ given above using the $SU(2)$ gauge field.
We denote the resulting theory by $U$.

Note that $\vec C$ is now a section of a nontrivial $\bR^4=\bC^2$ bundle over this $B^4$.
We would like $\vec C$ to tend to a constant, $\vec C\to \vec C_0$, close to the boundary $S^3$ of $B^4$.
We also need to preserve the $SU(2)_{k-1}$ symmetry.
According to the geometry explained in \eqref{mapping},
such a section behaves as $\vec C\propto x \vec C_0$ where $x\in S^3\simeq SU(2)$,
on the patch closer to the origin of $B^4$.
One explicit choice of such a section is simply to take $\vec C\propto \vec X$
near the origin, where $\vec X\colon B^4\to \bR^4$ is the obvious embedding.
According to \eqref{main-trick}, the $SU(2)_{k-1}$ symmetry is now the diagonal combination of
the $SU(2)_{k-1}$ symmetry of $T$, the $SU(2)_1$ symmetry of $\vec\psi$,
and the $SU(2)_{-1}$ symmetry of the four chiral multiplets $\vec X$.

By construction, this theory $U$ has a boundary $\nu(k)T'$ in the region $\vev{\Phi'}\to +\infty$,
where $\Phi'$ is the radial direction of $B^4$.
But the theory $U$ also has another boundary.
This is because the fiber of the theory $U$ at the origin $p\in B^4$ of the ball
is $T'(\vec C=0)$, which fails to be a compact theory in the region $\vev{\Phi}\to -\infty$.
To see what happens there,
note that the  superpotential of the theory $U$ in the vicinity of the origin $p$ has the form \begin{equation}
\propto \int d\theta^+ (\vec X \cdot \vec \Psi_- )f(\Phi).
\end{equation}
As $f(\Phi)$ is large and nonzero around $p$ in the region $\vev{\Phi}\to -\infty$, 
this gives large supersymmetric masses to $\vec X$ and $\vec \Psi_-$, removing them in the low energy limit,
leaving only $\Phi$ and $T$ in the system there. 
That is, what remains is the original theory $T$ trivially parameterized by $\Phi$ in the region $\vev{\Phi}\to -\infty$,
meaning that the theory $U$ has the theory $T$ as an incoming boundary.
This way we found the bordism $T \sim \nu(k)T'$.

\subsubsection{$T\sim \vec\psi T'$ when $\forget(T)\sim 0$}
Finally, suppose that an element $T\in \SQFT_{SU(2)}^{n,k}$ is null after forgetting $SU(2)$.
By the same trick, we can construct the following compact SQFT  $T'\in \SQFT^{n-4,k-1}_{SU(2)}$:
\begin{equation}
T':=\begin{tikzpicture}[scale=.7,baseline=(A)]
\begin{scope}
\clip (0,0) ellipse (3 and 1);
\shade[inner color=pink] (-3,0) ellipse (1 and 1) node {};
\end{scope}
\draw[thick] (0,1) arc[start angle=90, end angle=270, x radius=3, y radius= 1];
\draw[thick] (0,0) ellipse (1 and 1) node {$k$};
\node at (0,2.5)  {$T$};
\node (A) at (0,1.5)  {$\times$};

\draw[thick] (.5,2) arc[start angle=-90, end angle=90, x radius=3, y radius= .5];

\draw[->] (-4,-1.2) -- (+4,-1.2) node[below] {$\Phi$};
\node at (-1.5,-1.7) {$\underbrace{\phantom{aaaaaaaaa}}_{\displaystyle\overline{S}}$};
\node at (1.8,-1.7) {$\underbrace{\phantom{aaaaaaaaa}}_{\displaystyle S'}$};

\end{tikzpicture}\quad.
\end{equation}
That is, we use the assumed null bordism $\forget(T)\sim 0$ on the right hand side, i.e.~on the side of $S'$,
and we use the general null bordism $\nu(k) \forget(T)\sim0$ on the left hand side, i.e.~on the side of $\overline{S}$.
The resulting theory $T'$ is in $\SQFT_{SU(2)}^{n-4,k-1}$.

We now want to show that  $\vec\psi T' \sim T$.
For this, we  take the inspiration from the null bordism $\vec \psi\nu(k)\sim 0$ we discussed  in Sec.~\ref{psi-nu}.
Recall that the deformation used there was the superpotential \begin{equation}
\int d\theta^+ \vec X \cdot \vec \Psi_-,
\end{equation}
where $\vec X$ was the obvious embedding $\vec X: S^3\to \bR^4$.
Here we simply extend this map to $\vec X:B^4\to \bR^4$, 
which is again the obvious inclusion map.
This provides nonzero vacuum energy everywhere except at the origin of $B^4$.
Furthermore, it pairs up the fields $\vec X$ and $\vec \psi$ to make them massive,
and in the low energy limit only the theory $T$ remains.
This is what we wanted to show.

\section{Some consequences of the long exact sequence}
\label{sec:app}

Here we provide some applications of the long exact sequence \eqref{eq:seqx}.
Our first result is that $\SQFT^{n,-1}_{SU(2)}=0$ in general.
Next, we discuss how $\SQFT^{n,1}_{SU(2)}$ can be determined from the knowledge of $\SQFT^m$.
Finally, using these results, we study the chiral fermion spectrum of heterotic compactifications down to four dimensions with $SU(2)_1$  symmetry.

\subsection{On $\SQFT^{n,-1}_{SU(2)}$}
First, let us set $k=0$ in the long exact sequence, which is
\begin{equation}
\cdots\xrightarrow{\nu(0)\times} 
\SQFT_{SU(2)}^{n-4,-1}
\xrightarrow{\vec\psi \times}
\SQFT_{SU(2)}^{n,0}
\xrightarrow{\forget}
\SQFT^{n}
\xrightarrow{\nu(0)\times}
\SQFT_{SU(2)}^{n-3,-1}
\xrightarrow{\vec\psi\times}\cdots.
\end{equation}
We saw in Sec.~\ref{sec:concrete} that $\nu(0)\sim 0$.
Then $\SQFT^{n,0}_{SU(2)}\to \SQFT^n$ should be a surjection.
But we already know from our discussions in Sec.~\ref{second-lemma} that this is in fact an isomorphism, i.e.~\begin{equation}
\SQFT^{n,0}_{SU(2)}\simeq \SQFT^n.
\label{eq:isom}
\end{equation}
From the long exact sequence, we can then conclude that 
\begin{equation}
\SQFT^{n-4,-1}_{SU(2)}=0
\end{equation}
for any $n$. In words, any two-dimensional \Nequals{(0,1)} supersymmetric theory
with $SU(2)$ symmetry with level $-1$ (i.e.~on the right-moving, supersymmetric sector in our convention)
can be made to break supersymmetry spontaneously by a deformation.
This is a surprisingly general vanishing result, although the author does not know any immediate use of it.

\subsection{On $\SQFT^{n,1}_{SU(2)}$}

Next, let us set $k=1$, for which we have 
\begin{equation}
\cdots\xrightarrow{\nu(1)\times} 
\SQFT_{SU(2)}^{n-4,0}
\xrightarrow{\vec\psi \times}
\SQFT_{SU(2)}^{n,1}
\xrightarrow{\forget}
\SQFT^{n}
\xrightarrow{\nu(1)\times}
\SQFT_{SU(2)}^{n-3,0}
\xrightarrow{\vec\psi\times}\cdots.
\end{equation}
We use the isomorphism \eqref{eq:isom} again, to obtain the sequence \begin{equation}
\cdots\xrightarrow{\nu\times} 
\SQFT^{n-4}
\xrightarrow{\vec\psi \times}
\SQFT_{SU(2)}^{n,1}
\xrightarrow{\forget}
\SQFT^{n}
\xrightarrow{\nu\times}
\SQFT^{n-3}
\xrightarrow{\vec\psi\times}\cdots,
\label{main-long-exact-seq}
\end{equation}
where $\nu \in \SQFT^{-3}\simeq \SQFT^{-3,0}_{SU(2)}$ is the class of 
the $S^3$ sigma model with $\int_{S^3}H=1$.
Therefore we have the short exact sequence \begin{multline}
0\to (\Coker (\nu\times): \SQFT^{n-1} \to \SQFT^{n-4} ) \\
\xrightarrow{\vec\psi\times} 
\SQFT_{SU(2)}^{n,1}
\xrightarrow{\forget}\\
(\Ker (\nu\times): \SQFT^{n}\to \SQFT^{n-3})\to 0.
\label{short-exact-seq}
\end{multline}

This short exact sequence allows us to determine $\SQFT_{SU(2)}^{*,1}$
from the knowledge of $\SQFT^*$
and the action of $\nu\times$ on them,
up to an  ambiguity  inherent in the short exact sequence known as the extension problem.

Before proceeding, we mention that in the study of topological modular forms, there is a group
written as $(\TMF/\nu)^n$ which sits in a long exact sequence of the form \begin{equation}
\cdots\xrightarrow{\nu\times} 
\TMF^{n-4}
\xrightarrow{}
(\TMF/\nu)^{n-4}
\xrightarrow{}
\TMF^{n}
\xrightarrow{\nu\times}
\TMF^{n-3}
\xrightarrow{}\cdots.
\end{equation}
Assuming $\TMF^n\simeq  \SQFT^n$ and comparing with \eqref{main-long-exact-seq}, this would mean that 
\begin{equation}
(\TMF/\nu)^{n-4}\simeq \SQFT^{n,1}_{SU(2)}.
\label{tmfnu}
\end{equation}
As the groups $(\TMF/\nu)^n$ were computed in \cite{BrunerRognes},
this will allow us to obtain a lot of detailed information on $\SQFT^{n,1}_{SU(2)}$,
assuming the validity of the Stolz-Teichner proposal.
But we will not directly use the relation \eqref{tmfnu} in this paper.

\subsection{An application to heterotic compactifications to four dimensions}

Let us now use the short exact sequence \eqref{short-exact-seq} to answer a string theory question.
Suppose we consider heterotic compactifications down to four dimensions
with $SU(2)$ symmetry, where we further assume that the level is one.
Can it generate an odd number of chiral fermions in the doublet of $SU(2)$,
making the heterotic string compactification to have the Witten anomaly?

Note that such a compactification is described by a 2d \Nequals{(0,1)} superconformal field theory
of central charge $c_L=22$ and $c_R=9$, with $n=2(c_L-c_R)=26$.
Our strategy is to determine $\SQFT^{26,1}_{SU(2)}$
and  to extract the information of massless fermions from that.
We will perform the first step in Sec.~\ref{step1},
and study the second step in Sec.~\ref{step2}.

\subsubsection{Determination of $\SQFT^{26,1}_{SU(2)}$}
\label{step1}
As the short exact sequence \eqref{short-exact-seq} for $\SQFT^{26,1}_{SU(2)}$ depends on
$\SQFT^{25,22,26,23}$, we need to know them.
For this purpose, we need to rely on the conjecture that \begin{equation}
\SQFT^n \simeq \TMF^n.\label{ST}
\end{equation}
The structure of Abelian groups $\TMF^n$ is explained throughly in \cite{BrunerRognes}.
In particular, $\TMF^{25}$ and $\TMF^{26}$ are  known to vanish.
Then, we trivially have the conclusion that \begin{equation}
\TMF^{22}\simeq\SQFT^{22} \xrightarrow{\vec\psi\times} \SQFT^{26,1}_{SU(2)}
\end{equation} is actually an isomorphism.
In words, any 2d SQFT $T$ with gravitational anomaly $n=26$ with $SU(2)_1$  symmetry
is bordant to a theory which is given by an SQFT $T'$ with gravitational anomaly $n=22$ 
multiplied by a theory $\vec \psi$ of four Majorana-Weyl fermions,
such that the part $\vec\psi$ carries the entire $SU(2)_1$ symmetry.

Therefore, to understand $\SQFT^{26,1}_{SU(2)}$, we need to understand $\SQFT^{22}$.
For this purpose, the mod-2 elliptic genus discussed in \cite{Tachikawa:2023nne} is useful.
Given an SQFT $T'$ with gravitational anomaly $n\equiv -2$ mod 8,
its mod-2 elliptic genus $I^\text{mod2}_{T'}$ is defined by \begin{equation}
I^\text{mod2}_{T'} := \eta(q)^{-n}  \sum_a q^a \frac{\dim \cH_{H=P=a}}2   \in \bZ_2((q)).
\label{mod2}
\end{equation}
Here, $\bZ_2((q))$ is the ring of formal Laurent series in $q$ with $\bZ_2$ coefficients,
$\eta(q)$ is the Dedekind eta function,
$\cH$ is the Hilbert space of the theory $T'$ on an $S^1$ with R spin structure,
$H$ and $P$ are the Hamiltonian and the momentum operator satisfying $H-P\ge 0$ due to supersymmetry, 
and $\cH_{H=P=a}$ is the subspace where $H$ and $P$ have the common eigenvalue $a$.

On $\cH_{H=P=a}$, because of the commutation relation
$T(-1)^F=-(-1)^FT$ due to the gravitational anomaly $n\equiv -2$ mod 8, 
a state with $(-1)^F=+1$  and a state with $(-1)^F=-1$ always appear in pairs,
and therefore $\tr (-1)^F$ is automatically zero.
But this appearance in pairs means that $\dim\cH_{H=P=a}$ is always even,
and it can be shown that  $\dim \cH_{H=P=a}/2$ mod 2 is invariant under continuous deformation,
leading to the definition of the mod-2 elliptic genus \eqref{mod2} above.

Our normalization of $H$ and $P$ is that, 
for a conformal theory, we have $L_0-c_L/24=(H+P)/2$ and $\overline{L}_0-c_R/24=(H-P)/2$.
We note that the momentum eigenvalue $a$ has a shifted quantization $a\in \bZ+n/24$,
so that the $q$-expansion of the quantity \eqref{mod2} has integer  powers
and the mod-2 elliptic genus is indeed in  $\bZ_2((q))$.

Again by using the conjecture \eqref{ST} and the known mathematical facts concerning $\TMF^n$,
it was shown in \cite{Tachikawa:2023nne} that
$I^\text{mod2}_{T'}$ for $n\equiv -2$ mod 8 is necessarily a mod-2 reduction of 
an integral modular form of weight $-(n+2)/2$,
with a further condition that the mod-2 reduction of $\Delta(q)^\ell$ for $n=-24\ell-2$
does \emph{not} appear when $\ell\equiv 3,6,7$ mod 8,
where $\Delta(q)=\eta(q)^{24}$ is the modular discriminant. 

Our case $n=22$ corresponds to integral modular forms of weight $-12$, which is $\ell=-1$.
Therefore, among possible integral modular forms of weight $-12$, which are  integral linear combinations of \begin{equation}
  J(q)^m \Delta(q)^{-1} \qquad (m\ge 0),
\end{equation}
where $J(q)$ is the modular $J$-function,
the term $m=0$ does not arise. That is, the mod-2 elliptic genus of $T'$
is necessarily of the form 
 \begin{equation}
 I^\text{mod2}_{T'} = \sum_{m\ge 1} c_m  J(q)^m \Delta(q)^{-1}.
  \label{image}
\end{equation}
As $\Delta(q)=q+O(q^2)$ and $J(q)=q^{-1}+O(1)$,
we conclude that  $I^\text{mod2}_{T'}$, if nonzero, necessarily is of the form
$q^b(1+O(q))$ with $b\le -2$.

\subsubsection{Determination of the massless fermion spectrum}
\label{step2}

We would like to turn this knowledge concerning the theory $T'\in \SQFT^{22}$
to the knowledge concerning the spacetime fermion spectrum 
for the worldsheet theory $T=T'\vec\psi \in \SQFT^{26,1}_{SU(2)}$.
For this purpose it is useful to put the mod-2 elliptic genus \eqref{mod2}
into a broader context,
by generalizing it to  theories with an arbitrary gravitational anomaly $n$ and
an internal symmetry $G$ in the following manner.

\paragraph{KO-theoretic elliptic genus.}
For a theory $T\in \SQFT^{n,k}_G$, each subspace $\cH_{H=P=a}$
naturally determines a class $[\cH_{H=P=a}]$  of $\KO^{n}_G$, as was described in \cite[Appendix]{Tachikawa:2023nne}.\footnote{%
The $G$ action here can in general be twisted, giving rise to twisted equivariant KO-theory,
e.g.~when $G=\bZ_2$.
But for $G$ simple and simply-connected, it is known that no twists are involved.
}
We can then define \begin{equation}
I^\text{KO}_T := \eta(q)^{-n}  \sum_a q^a [\cH_{H=P=a}]   \in \KO^n_G((q)).
\end{equation}
The mod-2 elliptic genus \eqref{mod2} above is the special case for $\KO^n=\bZ_2$ when $n\equiv -1,-2$ mod 8,
and the ordinary elliptic genus is the special case for $\KO^n=\bZ$ when $n\equiv 0,4$ mod 8.
This generalized elliptic genus is  multiplicative, in the sense that \begin{equation}
I^\text{KO}_{T_1\times T_2}=I^\text{KO}_{T_1} \times I^\text{KO}_{T_2}.
\label{mul}
\end{equation}

\paragraph{The groups $\KO^n_G$ and the free fermion spectrum.}
At this point it is useful to note the following interpretation of the groups $\KO^{n}_G$.
Consider the set of Lorentz invariant free spin-1/2 fermion Lagrangians in $d$ dimensions
with symmetry $G$, up to isomorphism. 
This is naturally a monoid, where the addition is given by combining two Lagrangians.
We introduce an equivalence relation where we regard all  fermion Lagrangians which can be given nonzero masses as null.
The set of equivalence classes is now an Abelian group, since given a free fermion Lagrangian $\cL$, the Lagrangian $\cL+\overline{\cL}$, 
where $\overline{\cL}$ is the parity-reversed version of the theory $\cL$,
can always be made massive.
This group of equivalence classes of $G$-symmetric free fermion Lagrangians in $d$ dimensions is naturally isomorphic to $\KO^{d-2}_G$.
This statement is essentially equivalent to the celebrated periodic table of free topological superconductors \cite{Kitaev:2009mg,Ryu:2010zza} 
viewed from the boundary perspective,
and can be shown straightforwardly using the relation of spinor representations 
with Clifford modules, and the definition of $\KO^{d-2}_G$
in terms of Clifford modules in \cite{ABS}.

For example, take $G=\{e\}$ and $d=4$ or $8$. 
In this case, we know $\KO^{4-2}=0$ and $\KO^{8-2}=\bZ_2$.
As an example of free fermion Lagrangians, 
consider a single chiral spinor $\psi$, and 
the (would-be) mass term $m K^{\alpha\beta} \psi_\alpha \psi_\beta $,
where $K^{\alpha\beta}$ is the invariant tensor constructing a scalar 
out of two positive chirality spinors.
In four dimensions $K^{\alpha\beta}$  is antisymmetric,
and therefore the mass term is nonzero,
matching the fact that $\KO^{4-2}=0$.
In eight dimensions, $K^{\alpha\beta}$ is symmetric,
and therefore the would-be mass term is zero.
To write a mass term, we need \emph{two} positive chirality spinors $\psi$ and $\psi'$,
where $m K^{\alpha\beta} \psi_\alpha \psi'_\beta$ is now nonzero.
This matches the fact that $\KO^{8-2}=\bZ_2$.

Similarly, take $G=SU(2)$ and $d=4$. Then it is known that 
\begin{equation}
\KO^{4-2}_{SU(2)}= \bZ_2 V_2 \oplus \bZ_2 V_4 \oplus \cdots \oplus \bZ_2 V_{2n} \oplus \cdots,
\end{equation}
where $V_n$ is the class of the $n$-dimensional irreducible representation of $SU(2)$.
In terms of free fermion Lagrangians, 
this just means that a chiral fermion in an odd-dimensional irreducible representation 
of $SU(2)$ can be given a mass,
but to give a mass to chiral fermions in an even-dimensional irreducible representation,
we need an even number of them.

\paragraph{KO-theoretic elliptic genus and the massless fermion spectrum.}
This understanding of KO theory allows us to read off the chiral fermion spectrum in a heterotic compactification
using the internal worldsheet theory $T$ from its KO-theoretic elliptic genus $I^\text{KO}_T$.
To see this, 
recall that a compactification to spacetime dimensions $d$ requires 
an internal worldsheet superconformal field theory (SCFT)
with $(c_L,c_R)=(26-d,15-(3/2)d)$,
and that the massless fermions arise from the R-sector right-moving vacuum 
with $(H-P)/2=\overline{L}_0-c_R/24=0$.
Then 
the states with $L_0=0$ produce massless gravitinos and those with $L_0=1$ produce massless spin-1/2 fermions.
They have  $a=(H+P)/2=L_0-c_L/24=(d-26)/24$ and $(d-2)/24$, respectively.

Such a worldsheet SCFT has gravitational anomaly $n=22+d$.
If we expand $I^\text{KO}_T$ in the integer powers of $q$ as \begin{equation}
I^\text{KO}_T = \sum_{b} q^b v_b, \qquad v_b \in \KO^n_G((q)),
\end{equation}
the exponent $b$ is related to $H=P=a$ via $b=-n/24+a$.
Now, $\KO^n_G = \KO^{d+22}_G = \KO^{d-2}_G$ using the Bott periodicity,
and this is exactly what we need to read off the massless spectrum
of fermions in the spacetime theory.
Then the spectrum of the gravitinos and of the spin-1/2 fermions can be found
from the coefficients of $q^b$ of $I^\text{KO}_T$  at $b=-2$ and $b=-1$,
independent of $d$.

\paragraph{Heterotic compactifications to four dimensions with $SU(2)$ level 1 symmetry.}
Coming back to our original question of four-dimensional compactifications
with $SU(2)$ level 1 symmetry,
our theory $T$ is bordant to $T'\vec \psi$.
The R-sector vacuum of $\vec \psi$ consists of an $SU(2)$ doublet with $(-1)^F=+1$
and two singlets with $(-1)^F=-1$,
and has the value $H=P=4/24$. Therefore it has the elliptic genus \begin{equation}
I^\text{KO}_{\vec \psi} = (V_2-2V_1) + O(q) \in \KO^4_{SU(2)} ((q)),
\end{equation}
where $\KO^4_G=\mathop{\mathrm{KSp}}\nolimits^0_G$ is the 
Grothendiek group of quaternionic representations of $G$, which for $G=SU(2)$ is \begin{equation}
\KO^4_{SU(2)}= 2\bZ V_1 \oplus \bZ V_2 \oplus 2\bZ V_3 \oplus \cdots.
\end{equation}

Combining this with \eqref{image} and \eqref{mul}, we conclude that $
I^\text{KO}_T,
$ 
if nonzero, necessarily has a leading term \begin{equation}
q^b(V_2 + O(q)) \in \KO^{4-2}_{SU(2)}((q))
\end{equation}
with $b\le -2$.
This is the  generic situation for a general SQFT with $n=26$ with $SU(2)$ level 1 symmetry.

But for a heterotic compactification, we need an SCFT with $(c_L,c_R)=(22,9)$,
for which $b=L_0-2$ and therefore $b\ge -2$ via unitarity.
The only possibility of  nonzero $I^\text{KO}_T$ for such an SCFT is then $b=-2$,
which would correspond to having an odd number of doublets at $L_0=0$.
This is actually disallowed again by the unitarity of this SCFT, 
since the $SU(2)$ level 1 symmetry
dictates that any primary in the doublet of $SU(2)$ should have $L_0\ge 1/4$.\footnote{%
In spacetime terms, this is the standard fact that in Minkowski supergravity,
gravitinos are never charged under continuous gauge fields.}
We conclude that, 
for a heterotic compactification to four dimensions with $SU(2)$ level 1 symmetry,
$I^\text{KO}_T=0$.
This implies that the spacetime fermions contain every even-dimensional irreducible representations of $SU(2)$ even times.
This is a stronger result than the absence of the Witten anomaly, 
which only requires the sum of the number of irreducible representations
of dimension $2+4k$ is even.

\section*{Acknowledgements}

The author thanks Y.-H. Lin and M. Yamashita for explaining the content of \cite{LinYamashita} to him
and for urging him to come up with an SQFT derivation of their long exact sequences.
He also thanks Y.-H. Lin for carefully reading an earlier version of the paper and giving many insightful comments,
and thanks Yanming Su and Yi Zhang for similarly providing various useful comments on the draft.

The author also owes a lot to  Y. Sato and T. Watari, who asked the author 
during an IPMU tea time on a cold day  just before the corona virus pandemic in early 2020
a prophetic question of
whether the $SU(2)$ Witten anomaly was absent in arbitrary heterotic compactifications
down to four dimensions.
This led the author to this whole endeavor connecting topological modular forms
and two-dimensional \Nequals{(0,1)} SQFTs in the last five years.
This paper is an important milestone to the author, where $\SQFT^{26,1}_{SU(2)}$ was finally computed,
which was one of the earliest goals in this project.

The author is supported in part  
by WPI Initiative, MEXT, Japan at Kavli IPMU, the University of Tokyo
and by JSPS KAKENHI Grant-in-Aid (Kiban-C), No.24K06883. 

\def\arxivfont{\rm}
\bibliographystyle{ytamsalpha}
\let\originalthebibliography\thebibliography
\renewcommand\thebibliography[1]{
  \originalthebibliography{#1}
  \setlength{\itemsep}{2pt plus 0.4ex}
}
\bibliography{ref}

\end{document}